\documentstyle[11pt]{article}
\newcommand{\be}[1]{\begin{equation}\label{#1}}
\newcommand{\ba}[1]{\begin{eqnarray}\label{#1}}
\newcommand{\ee}{\end{equation}}
\newcommand{\ea}{\end{eqnarray}}
\newcommand{\non}{\nonumber\\\rule{0pt}{30pt}}

\newcommand{\dis}{\displaystyle}
\newcommand{\eq}[1]{(\ref{#1})}

\begin{document}
\begin{flushright}
LPENSL-TH-02/02\\
\end{flushright}
\par \vskip .1in \noindent

\vspace{24pt}

\begin{center}
\begin{LARGE}
{\bf Emptiness formation probability of the $XXZ$
spin-$\textstyle{\frac{1}{2}}$ Heisenberg chain at
$\Delta=\textstyle{\frac{1}{2}}$}
\end{LARGE}

\vspace{50pt}

\begin{large}

{\bf N.~Kitanine}\footnote[1]{Graduate School of Mathematical
Sciences, University of Tokyo, Japan,
kitanine@ms.u-tokyo.ac.jp\par
\hspace{2mm} On leave of absence from Steklov Institute at
St. Petersburg, Russia},~~
{\bf J.~M.~Maillet}\footnote[2]{ Laboratoire de Physique, UMR 5672 du CNRS,
ENS Lyon,  France,
 maillet@ens-lyon.fr},~~
{\bf N.~A.~Slavnov}\footnote[3]{ Steklov Mathematical Institute,
Moscow, Russia, nslavnov@mi.ras.ru},~~
{\bf V.~Terras}\footnote[4]{Department of Physics and Astronomy,
Rutgers University, USA, vterras@physics.rutgers.edu \par
\hspace{2mm} On leave of absence from LPMT, UMR 5825 du CNRS,
Montpellier, France}
\end{large}

\vspace{70pt}

\centerline{\bf Abstract} \vspace{1cm}
\parbox{12cm}{\small Using a  multiple integral representation for
the correlation functions, we compute the emptiness formation
probability of the $XXZ$ spin-$\frac{1}{2}$ Heisenberg chain at
anisotropy $\Delta=\textstyle{\frac{1}{2}}$. We prove it is
expressed in term of the number of alternating sign matrices.}
\end{center}

\vspace{30pt}

The Hamiltonian of the $XXZ$ spin-$1 \over 2$ Heisenberg chain
is given by
\be{IHamXXZ} H=\sum_{m=1}^{M}\left(
\sigma^x_{m}\sigma^x_{m+1}+\sigma^y_{m}\sigma^y_{m+1}
+\Delta(\sigma^z_{m}\sigma^z_{m+1}-1)\right).
\ee
Here $\Delta$ is the anisotropy parameter, $\sigma^{x,y,z}_{m}$
denote the usual Pauli matrices acting on the quantum space at
site $m$ of the chain. The emptiness formation probability
$\tau(m)$ (the probability to find in the ground state a
ferromagnetic string of length $m$) is defined as the following
expectation value
\be{EMPtau}
\tau(m)=\langle\psi_g|\prod_{k=1}^m\frac{1-\sigma_k^z}2
|\psi_g\rangle,
\ee
where $|\psi_g\rangle$ denotes the normalized ground state. In the
thermodynamic limit ($M\to\infty$), this quantity can be expressed
as a multiple integral with $m$ integrations
\cite{JimMMN92,JimM96, JimML95,KitMT00}. Recently, in the article
\cite{KitMST02}, a new multiple integral representation for
$\tau(m)$ was obtained; for $\Delta=\cos\zeta$, $0<\zeta<\pi$, one
has
\be{tauhm} \tau(m)=\lim_{\xi_1,\dots\xi_m\to-\frac{i\zeta}2}
\tau(m,\{\xi_j\}), \ee
where
\be{EFPtausym} \tau(m,\{\xi_j\})=\frac1{m!}
\int\limits_{-\infty}^{\infty}\frac{Z_m(\{\lambda\},\{\xi\})}
{\prod\limits_{a<b}^m \sinh(\xi_a-\xi_b)}
{\det}_m\left(\frac{i}{2\zeta
\sinh\frac{\pi}{\zeta}(\lambda_j-\xi_k)}\right) \,d^m\lambda, \ee
with
\be{EFPZm}
Z_m(\{\lambda\},\{\xi\})=
\prod\limits_{a=1}^{m}\prod\limits_{b=1}^{m}
\frac{\sinh(\lambda_a-\xi_b)\sinh(\lambda_a-\xi_b-i\zeta)}
{\sinh(\lambda_a-\lambda_b-i\zeta)}
\cdot\frac{{\det}_m\left(\frac{-i\sin\zeta}
{\sinh(\lambda_j-\xi_k)\sinh(\lambda_j-\xi_k-i\zeta)}\right)}
{\prod\limits_{a>b}^m\sinh(\xi_a-\xi_b)}.
\ee

In this letter, we consider the particular case
$\Delta=\frac{1}{2}$ ($\zeta=\pi/3$). Recently several interesting
conjectures were obtained for the ground state of the model at
this special value of the anisotropy parameter $\Delta$
\cite{Str01,RazS01,BatdeGN01,deGNM01}. Note that the unitary
transformation $UH_{\Delta} U^{-1}=-H_{-\Delta}$,
$U=\prod_{j=1}^{M \over 2}\sigma_{2j}^z$ relates our Hamiltonian
\eq{IHamXXZ} for $\Delta=\frac{1}{2}$ to the case
$\Delta=-\frac{1}{2}$ in \cite{RazS01}. In particular, it was
conjectured in \cite{RazS01} that, in this case, the emptiness
formation probability is equal to
\be{Str}
\tau(m)=\left(\frac{\sqrt3}2\right)^{3m^2}
\prod_{k=1}^m\frac{\Gamma(k-\frac13)\Gamma(k+\frac13)}
{\Gamma(k-\frac12)\Gamma(k+\frac12)}.
\ee
The aim of this letter is to give the proof of this conjecture
using the representations \eq{tauhm}--\eq{EFPZm}.

Observe first that for $\zeta=\pi/3$,
\ba{EFPZm1} &&{\dis\hspace{-10mm} Z_m(\{\lambda\},\{\xi\})=
\frac{(-1)^{\frac{m^2-m}{2}}}{2^{m^2+m}}
\prod_{a>b}^{m}\frac{\sinh3(\xi_b-\xi_a)}
{\sinh(\xi_b-\xi_a)\sinh(\xi_a-\xi_b)}}\non
&&{\dis\hspace{8mm}
\times{\det}_m\left(\frac{1}
{\sinh(\lambda_j-\xi_k)\sinh(\lambda_j-\xi_k-i\zeta)}\right)
\frac{{\det}_m\left(\frac{1}{\sinh(\lambda_j-\xi_k
+\frac{i\pi}{3})}\right)}
{{\det}_m\left(\frac{1}{\sinh3(\lambda_j-\xi_k)}\right)}.}
\ea
Here we have used the identities,
\be{SZCauchy}
{\det}_n\frac1{\sinh(x_j-y_k)}=
\frac{\prod\limits_{a>b}^n\sinh(x_j-x_k)\sinh(y_k-y_j)}
{\prod\limits_{a,b=1}^n\sinh(x_j-y_k)},
\ee
and $\sinh(3x)=4\sinh(x)\sinh(x+i\pi/3)\sinh(x-i\pi/3)$.
Substituting \eq{EFPZm1} into \eq{EFPtausym}, we obtain
\ba{tau} &&{\dis\hspace{-1mm}
\tau(m,\{\xi_j\})=\left(\frac{3i}{4\pi}\right)^m
\frac{(-1)^{\frac{m^2-m}{2}}}{2^{m^2}m!}
\prod_{a>b}^{m}\frac{\sinh3(\xi_b-\xi_a)}{\sinh(\xi_b-\xi_a)}
\prod\limits_{a,b=1\atop{a\ne b}}^{m}\sinh^{-1}(\xi_a-\xi_b)}\non
&&{\dis\hspace{-1mm}
\times\int\limits_{-\infty}^\infty\,d^m\lambda\quad
{\det}_m\left(\frac{1}{\sinh(\lambda_j-\xi_k+\frac{i\pi}3)}\right)
{\det}_m\left(\frac{1}{\sinh(\lambda_j-\xi_k)
\sinh(\lambda_j-\xi_k-\frac{i\pi}3)}\right).}
\ea
Due to the symmetry properties of the integrand,  we can replace
the first determinant with the product of its diagonal elements
multiplied by $m!$. Then, we insert each of  these diagonal
elements into the corresponding line of the second determinant. By
this procedure, the integrals over the variables $\lambda$ are
decoupled and we can integrate each line of the determinant
separately. Let us set $\xi_k=\varepsilon_k-i\pi/6$. We obtain
\ba{tau1} &&{\dis\hspace{1mm}
\tau(m,\{\varepsilon_j\})=(-1)^{\frac{m^2-m}{2}} 3^{m}2^{-m^2}
\prod_{a>b}^{m}\frac{\sinh3(\varepsilon_b-\varepsilon_a)}
{\sinh(\varepsilon_b-\varepsilon_a)} \prod\limits_{a,b=1\atop{a\ne
b}}^{m} \frac{1} {\sinh(\varepsilon_a-\varepsilon_b)}}\non
&&{\dis\hspace{10mm}
\times\quad
{\det}_m\left(\int\limits_{-\infty}^\infty
\frac{d\lambda}{4\pi\cosh(\lambda-\varepsilon_j)
\sinh(\lambda-\varepsilon_k-\frac{i\pi}6)
\sinh(\lambda-\varepsilon_k+\frac{i\pi}6)}
\right).}
\ea
The computation of the integral over $\lambda$ in \eq{tau1} leads
to
\be{tau2}
\tau(m,\{\varepsilon_j\})=\frac{(-1)^{\frac{m^2-m}{2}}}
{2^{m^2}}\prod_{a>b}^{m}
\frac{\sinh3(\varepsilon_b-\varepsilon_a)}
{\sinh(\varepsilon_b-\varepsilon_a)} \prod\limits_{a,b=1\atop{a\ne
b}}^{m}\frac{1}{ \sinh(\varepsilon_a-\varepsilon_b)}\cdot
{\det}_m\left(\frac{3\sinh\frac{\varepsilon_j-\varepsilon_k}{2}}
{\sinh\frac{3(\varepsilon_j-\varepsilon_k)}{2}}\right). \ee
To obtain the emptiness formation probability \eq{EMPtau}, one has
to take the homogeneous limit $\varepsilon_j\to0$. Using the fact
that
\be{lim} \lim_{x_j\to x\atop{y_k\to y}} \frac{{\det}_m f(x_j-y_k)}
{\prod\limits_{a>b}^{m}(x_a-x_b)(y_b-y_a)}
=\prod_{n=0}^{m-1}(n!)^{-2}
{\det}_m\Bigl[f^{(j+k-2)}(z)\Bigr],\qquad z=x-y,
\ee
we finally obtain
\be{taures}
\tau(m)=(-1)^{\frac{m^2-m}{2}}
3^{\frac{m^2+m}2}2^{-m^2}
\prod_{n=0}^{m-1}(n!)^{-2}
{\det}_m\left[\frac{\partial^{j+k-2}}
{\partial x^{j+k-2}}\frac{\sinh\frac{x}{2}}
{\sinh\frac{3x}{2}}\right]_{x=0}.
\ee
The determinant in \eq{taures} can be computed using the following
identity \cite{Kup97},
\be{mainf}
\frac{1}{\prod\limits_{j>k}^{m}\sinh^2\beta(j-k)}
\cdot{\det}_m\frac{\sinh\alpha(j+k-1)}{\sinh\beta(j+k-1)}=
2^{m^2-m}\prod_{j=1}^{m}\prod_{k=1}^{m}
\frac{\sinh(\alpha+\beta(j-k))}{\sinh\beta(j+k-1)},
\ee
(see \cite{Kup97} for the proof). The determinant \eq{taures} is a
particular case of \eq{mainf}. Indeed, we can consider the case
$\beta=3\alpha$, $\alpha\to 0$, and apply \eq{lim} for $x_j=\alpha
j$, $y_k=\alpha(1-k)$. Then,
\ba{detres}
&&{\dis\hspace{-10mm}
\prod_{n=0}^{m-1}(n!)^{-2}
{\det}_m\left[\frac{\partial^{j+k-2}}
{\partial x^{j+k-2}}\frac{\sinh\frac{x}{2}}
{\sinh\frac{3x}{2}}\right]_{x=0}
=3^{m^2-m}\prod_{j=1}^{m}\prod_{k=1}^{m}
\frac{j-k+\frac{1}{3}}{j+k-1}}\non
&&{\dis\hspace{20mm} =(-1)^{\frac{m^2-m}{2}}3^{-\frac{m+m^2}{2}}
\prod_{k=0}^{m-1}\frac{(3k+1)!}{(m+k)!}.} \ea
Substituting these expressions into \eq{taures}, we finally obtain
\be{mainres}
\tau(m)=\left(\frac{1}{2}\right)^{m^2}
\prod_{k=0}^{m-1}\frac{(3k+1)!}{(m+k)!}.
\ee
Observe that the quantity $A_m=\prod_{k=0}^{m-1}(3k+1)!/(m+k)!$ is
the number of alternating sign matrices of size $m$ \cite{Zei94}.
Using
\be{3gam} \Gamma(3z)=\frac{1}{2\pi}3^{3z-1/2}
\Gamma(z)\Gamma(z+1/3)\Gamma(z+2/3),
\qquad\Gamma(k+1/2)=\frac{\sqrt\pi}{2^k}(2k-1)!!, \ee
one can easily check the equivalency of \eq{mainres} and \eq{Str}.
Thus \eq{Str} is proved.

The asymptotic behavior of $\tau(m)$ for $m\to\infty$ can be
evaluated using the Stirling formula \cite{RazS01}:
\be{asytau}
\tau(m)\to c\left(\frac{\sqrt3}2\right)^{3m^2}
m^{-\frac5{36}},\qquad m\to\infty,
\ee
with
\be{c}
c=\exp\left[\int_0^\infty\left(\frac{5e^{-t}}{36}
-\frac{\sinh\frac{5t}{12}\sinh\frac{t}{12}}
{\sinh^2\frac{t}{2}}\right)\frac{dt}t\right].
\ee
\vspace{1cm}

\section*{Acknowledgments}

N. K. would like to thank the University of York, the SPhT in
Saclay, JSPS and the Tokyo University  for financial support. N.
S. is supported by the grants INTAS-99-1782, RFBR-99-01-00151,
Leading Scientific Schools 00-15-96046, the Program Nonlinear
Dynamics and Solitons and by CNRS. J.M. M. is supported by CNRS.
V. T is supported by DOE grant DE-FG02-96ER40959 and by CNRS.  N.
K, N. S. and V. T. would like to thank the Theoretical Physics
group of the Laboratory of Physics at ENS Lyon for hospitality,
which makes this collaboration possible.

\end{document}